\documentclass[aps,twocolumn,groupedaddress]{revtex4}
\usepackage{amsmath,amssymb}

\begin{document}
\bibliographystyle{apsrev}


\title{Markovian diffusion approach to the relativistic ideal gas system}




\author{Ryszard Zygad\l{}o }
\affiliation{Marian Smoluchowski
Institute of Physics, Jagiellonian University, Reymonta 4, PL--30059 Krak\'ow,
Poland}


\date{\today}

\begin{abstract}
It is shown that the usual identification of
certain averages with the relativistic thermodynamical
functions is not possible in the moving reference frame description.
The Brownian motion approach is used  as the Markovian
kinetic description of the relativistic ideal gas system.
\end{abstract}

\maketitle


The number of  phenomena,
the  {\it critical slowing down} [1], {\it noise--induced transitions} [2],
{\it stochastic resonance} [3], {\it ratchet dynamics} [4], and
{\it resonant activation} [5], have been discovered by  studying
the effect of the noise in nonlinear systems [6].
The mechanical model for the most of applications is the
(Newtonian) Brownian particle in a certain external potential. In spite
of the significant progress with understanding the nonlinear stochastic behavior
the rigorous results, going beyond the single-event stationary description,
are not available for multi-variable systems and thus usually the overdamped kinetic
is considered. The stochastic description is then provided by the first order
Langevin equation or, in the case of ({\it thermal}) Gaussian white noise
(GWN), by the corresponding (Smoluchowski--) Fokker--Planck equation
[7] in the {\it configuration} space.
Note that the pioneering works [8] on the free Brownian motion also deal with
the diffusion, but in the {\it momentum} space,
\begin{equation}            \dot{p}_t= -\gamma v +\xi_t,
  \end{equation}
where the kinematic friction is proportional to the
velocity $v = p/m$ and
$\langle\xi_t \xi_s \rangle =2D\delta(t-s)$. The Eq.~(1)
leads to the correct stationary state of thermal equilibrium
\begin{equation}
W(p) \propto e^{-(\gamma/D) p^2/2m} = e^{-\beta \varepsilon}
\end{equation}
of an ideal gas system
and provides the fluctuation--dissipation relation $\gamma/D=\beta$
(or $D=\gamma k_B T$). The first order Langevin equation in the momentum
space may be thus considered as a diffusion Markovian model of
thermal equilibration for spatially homogeneous free particles systems.
The main idea of the present paper is to apply the same approach to
the ideal gas of relativistic particles. The relation between momentum
and velocity
$v= d\varepsilon/dp =p/\sqrt{p^2 +m^2}$, where $\varepsilon = \sqrt{p^2 +m^2}$
is the energy of the particle, is then nonlinear, as well as the kinetic equation
\begin{equation}            \dot{p}_t= -\gamma p/\sqrt{p^2 +m^2} +\xi_t.
\end{equation}
The Gibbs--Boltzmann 1D normalized stationary distribution easily follows
from Fokker--Planck theory [7]
\begin{equation}
W(p) = {1 \over 2m K_1(m\gamma/D)} e^{-(\gamma/D) \sqrt{p^2 +m^2}}
\end{equation}
and the Einstein relation $\gamma/D=\beta$ remains unchanged.
$K_1 (m\beta)$ is a modified Bessel function.

The use of the diffusion description in thermodynamics is related
to the Markovian approximation of the kinetic theory. The general
Kramers--Fokker--Planck kinetic equation in the {\it phase} space,
see e.g.\ Ref.~[9], simplifies to the form
\begin{equation}
\partial_t P(\vec{p},t) =
\gamma \nabla \circ (\vec{p} P/\varepsilon) +
D \nabla^2 P
\end{equation}
for a spatially homogeneous 3D system of a free relativistic
particles with a ({\it reduced}) potential interaction
with the thermal bath. The potential interaction
corresponds to the state-independent
diffusion at the level of equation (5).
$\nabla = \partial/ \partial \vec{p}$ and $\varepsilon =
\sqrt{\vec{p}^2 +m^2}$. The 3D generalization of the Langevin
description (3) is obvious, however in contrast to the Newtonian
case it results with the set of three {\it coupled} equations with
additive independent noises. The second remark concerns the specific
requirements of the special relativity theory. Note that the
Gibbs--Boltzmann equilibrium distribution is established in the
{\it distinguished} resting reference frame of the thermal bath
(and of the considered system as a whole),
in which the system and the bath remain in contact and in
equilibrium. The parameter $\beta = 1/k_B T$ is given by the
temperature of the thermal bath [10].
The any corresponding probability density distributions in
a momentum space,
$W(\vec{p})$ and $W^\prime (\vec{p}^\prime)$
in a resting and moving (along $x$-axis) with a constant velocity $V$
reference frame,
respectively,
are related by [11]
\begin{equation}
\varepsilon W(\vec{p}) = \varepsilon^\prime W^\prime (\vec{p}^\prime),
\end{equation}
where $\varepsilon$ and $\vec{p}$  should be expressed as  functions of
$\varepsilon^\prime$ and $\vec{p}^\prime$, according to
the Lorentz transformation
\begin{equation}
\varepsilon= (\varepsilon^\prime + V p_x^\prime)/\sqrt{1-V^2},
\end{equation}
\begin{equation}
p_x= (p_x^\prime + V\varepsilon^\prime)/\sqrt{1-V^2}, \quad p_y =p_y^\prime,
\quad p_z= p_z^\prime.
\end{equation}
Because the time increments, counted from the initial
preparation of the system, are given by $t =t^\prime /\sqrt{1-V^2}$ and $t^\prime$,
respectively, the Eq.~(6) leads to following relation
\begin{equation}
P^\prime (\vec{p}^\prime, t^\prime) =
{1+ Vp_x^\prime/\varepsilon^\prime \over \sqrt{1-V^2} } P \Bigl({p_x^\prime+
V\varepsilon^\prime \over \sqrt{1-V^2}},\, p_y^\prime, \, p_z^\prime, \,
{t^\prime \over \sqrt{1-V^2}} \Bigr)
\end{equation}
between an arbitrary $P(\vec{p},t)$ and the corresponding
$P^\prime (\vec{p}^\prime, t^\prime)$. Because the stochastic evolution
is Markovian the process is completely determined in both reference frames
if the time dependent single-event probability density $P(\vec{p},t)$ is
known. The resulting from Eqs.~(5) and (9) Fokker--Planck equation in
a moving reference frame is however very complicated in 3D, so
the next consideration are restricted to one dimensional case.

The (1D) equilibrium distribution in a moving reference frame results
directly from  Eqs.~(4), (6), (7), and (8),
\begin{equation}
W^\prime (p^\prime ) = \! { 1+V p^\prime /\sqrt{{p^\prime}^2 +m^2} \over
2m\sqrt{1-V^2} K_1(m\beta)}
\exp\Bigl[-{\sqrt{{p^\prime}^2 +m^2} + V p^\prime
\over \beta^{-1} \sqrt{1-V^2}} \Bigr],
\end{equation}
and thus it is not involved with particular kinetic model.
From the viewpoint of the probability theory the difference
between distribution (10)
and the Boltzmann distribution (4)
is {\it essential},
i.e., it cannot be regarded by the change of
the  {\it location parameters} [12], in contrast
to the Newtonian case.
The number of averages may by computed
from the generic formula [13]
\begin{equation}
\int_{-\infty}^{+\infty} dx {e^{-a\sqrt{x^2 + m^2} -bx}\over \sqrt{x^2 + m^2}}
  = 2 K_0 (m\sqrt{a^2 - b^2}),
\end{equation}
where $a > |b|$, by the successive differentiation with
respect to $a$ and/or $b$.
\begin{equation}
\langle \varepsilon^\prime \rangle =
{- m K_1^\prime (m\beta)  \over \sqrt{1-V^2}K_1 (m\beta)},
\end{equation}
\begin{equation}
\langle p^\prime \rangle =
{mV K_1^\prime (m\beta)  \over \sqrt{1-V^2} K_1 (m\beta)},
\end{equation}
\begin{equation}
\langle {\varepsilon^\prime}^2 \rangle =
{1  \over \beta^2 (1-V^2)K_1 }
[m^2 \beta^2 K_1^{\prime\prime} - m\beta V^2 {K_1^\prime} + V^2 K_1],
\end{equation}
\begin{equation}
\langle {p^\prime}^2 \rangle =
{1  \over \beta^2 (1-V^2) K_1 }
[m^2\beta^2 V^2 K_1^{\prime\prime} - m\beta {K_1^\prime} + K_1],
\end{equation}
\begin{equation}
\langle {p^\prime}\varepsilon^\prime \rangle =
{V  \over \beta^2 (1-V^2)K_1 }
[m^2\beta^2 K_1^{\prime\prime} - m\beta {K_1^\prime} + K_1].
\end{equation}
The omitted arguments of (14--16) are equal $m\beta$. The
averages in the resting reference frame
are obtained by putting
$V=0$ in Eqs.~(12--16).

The stationary averages of twovectors
$\langle (\varepsilon, \, p) \rangle = (-mK_1^\prime/K_1, \, 0)$ and
$\langle (\varepsilon^\prime, \,  p^\prime) \rangle =
(\langle \varepsilon\rangle /\sqrt{1-V^2}, \,
-V\langle \varepsilon\rangle /\sqrt{1-V^2})$,
in the resting and moving reference frame satisfy
the Eqs.~(7, 8). We have also
$\langle p^\prime \rangle = -V \langle \varepsilon^\prime \rangle$.
Thus the equilibrium mean energy and mean momentum of the
Brownian particle may be considered as the components of appropriate twovector
of a free particle with {\it renormalized resting mass}
\begin{equation}
m(T)=-mK_1^\prime(m\beta)/K_1(m\beta)\,\, [=\langle \varepsilon \rangle].
\end{equation}

The Eqs.~(14) and (15) show that the invariant relation
$\langle {\varepsilon^\prime}^2 -{p^\prime}^2 \rangle =m^2$
holds identically, because the equation
defining the Bessel function $K_1 (m\beta)$,
\begin{equation}
(m\beta)^2 K_1^{\prime\prime} + (m\beta) {K_1^\prime} - [1+(m\beta)^2]  K_1 =0,
\end{equation}
is obtained as the condition.

The single collision with the certain ``wall" of 1D box
[of a volume (length) ${\cal V}$] in the resting reference frame is
associated with energy-momentum transfer $(\varepsilon, \, p)
-(\varepsilon, \, -p)=(0, \, 2p)$ and the period between successive events
is $2{\cal V}/v = 2\varepsilon {\cal V}/p $.
Thus the  pressure is
\begin{equation}
{\cal P} = \langle p^2/({\cal V}\sqrt{p^2+m^2})\rangle =
1/\beta {\cal V} = k_B T /{\cal V}.
\end{equation}
The corresponding momentum transfer in the moving
reference frame follows from the Lorentz transformation of
$(0,\, 2p)$ and it is equal $2p/\sqrt{1-V^2}$. Simultaneously the period
changes according to the Lorentz dilatation, so finally [14]
\begin{equation}
{\cal P}^\prime = {\cal P}/(1-V^2) = k_B T/({\cal V}^\prime \sqrt{1-V^2}).
\end{equation}
The (thermodynamical) entropy is defined in the resting reference frame by
\begin{eqnarray}
&S&=-k_B \langle \ln [h W(p)/{\cal V}] \rangle  \nonumber \\ && =
k_B \{\beta \langle \varepsilon \rangle + \ln [2m {\cal V}
K_1(m\beta)/h]\},
\end{eqnarray}
where the Planck constant is introduced for dimensional purposes.
Using  $U^\prime = \langle \varepsilon^\prime \rangle = U/\sqrt{1-V^2}$
and
\begin{equation}
dU^\prime = T^\prime dS^\prime -{\cal P}^\prime d{\cal V}^\prime
=(TdS -{\cal P} d{\cal V})/\sqrt{1-V^2},
\end{equation}
in view of Eq.~(20) one obtains the relation
$T^\prime dS^\prime
=TdS/\sqrt{1-V^2}$. Assuming $T^\prime = \alpha(V) T$ and using
Eq.~(21) the relations [14]
\begin{equation}
T^\prime = T, \quad {\rm and} \quad S^\prime = S/\sqrt{1-V^2},
\end{equation}
supporting the Landsberg choice of the invariant temperature
[15] are obtained.

The identification of averages (19) and (21) with thermodynamical
functions is not possible in a moving reference frame.
Neither the quantity  defined by
$-k_B \langle \ln [h W^\prime(p^\prime)/{\cal V^\prime}] \rangle$
is equal to the entropy (23)
$S^\prime$, nor the quantity $
 \langle {p^\prime}^2/
({\cal V}^\prime \sqrt{{p^\prime}^2+m^2}) \rangle$  is equal to
the pressure (20) ${\cal P}^\prime$, which follows
from {\it explicit} expressions.
For the same reasons the relation between the
energy fluctuation and the heat capacity
\begin{equation}
{\rm Var}(\varepsilon)
= -\partial \langle \varepsilon \rangle/\partial \beta,
\end{equation}
do not hold in the moving reference frame,
\begin{equation}
{\rm Var}(\varepsilon^\prime) =
{m^2 K_1^{\prime\prime} - {m V^2 \over \beta} {K_1^\prime} + {V^2 \over \beta^2}
K_1 -
(m K_1^\prime)^2/K_1
  \over (1-V^2)K_1 },
\end{equation}
and
\begin{equation}
-{\partial \langle \varepsilon^\prime \rangle \over \partial \beta}
= {m^2 K_1^{\prime\prime} -(m K_1^\prime)^2/K_1 \over \sqrt{1-V^2} K_1 }.
\end{equation}
The r.h.s. of Eqs.~(25, 26) are different, if $V \ne 0$.

The Fokker--Planck equation of the
transformed  (1D) kinetic, Eqs.~(5) and (9), reads
\begin{eqnarray}
{\partial \over \partial t^\prime} P^\prime (p^\prime,\, t^\prime) && =
- \gamma{\partial \over \partial p^\prime}
{(v^\prime + V)  \over (1+V v^\prime)^2} P^\prime  \\
&&+D  \Bigl[ {\partial \over \partial p^\prime}{(1-V^2)^{1/4} \over 1+Vv^\prime}
{\partial \over \partial p^\prime}{(1-V^2)^{1/4} \over 1+Vv^\prime} \Bigr]
P^\prime \nonumber
\end{eqnarray}
and thus corresponds to the following multiplicative
\begin{equation}
{dp^\prime \over dt^\prime}=-\gamma {v^\prime + V \over (1+V v^\prime)^2}
+ {(1-V^2)^{1/4} \over 1+Vv^\prime} {\xi_{t^\prime}},
\end{equation}
Langevin equation (written in the Stratonovich interpretation [7]),
where $v^\prime = p^\prime/\sqrt{{p^\prime}^2 +m^2}$.
The Eq.~(28) may be obtained directly by appropriate transformation
of stochastic processes entering to Eq.~(3). We have
$d{p}_t =  [1 + (p^\prime /\varepsilon^\prime)] dp^\prime/\sqrt{1-V^2}$
and $dt = dt^\prime/\sqrt{1-V^2} $, so
\begin{equation}
d{p}_t/ dt =
(1+V v^\prime) dp^\prime/dt^\prime.
\end{equation}
Dividing the sides of (8) by the respective sides of (7) one has
\begin{equation}
v= (v^\prime + V)/(1+V v^\prime),
\end{equation}
which is the relativistic rule of collecting the velocities.
Because $\langle \xi_{t} \xi_0 \rangle = 2D \delta (t) =
2D\delta (t^\prime/\sqrt{1-V^2}) = \sqrt{1-V^2}
[2D\delta(t^\prime)]$ the GWN transforms according
\begin{equation}
 \xi_t =(1-V^2)^{1/4} \xi_{t^\prime},
\end{equation}
where ${\xi_{t^\prime}}$ is again a
white Gaussian noise with the same parameter $D$.
Collecting (29--31) we obtain the Langevin equation (28)
in Stratonovich interpretation (because the ordinary calculus
has been consequently used [7]).
The Eq.~(28) is a rather complicated nonlinear stochastic equation
with (nonlinearly coupled) multiplicative (i.e., the state dependent) noise.
For the special
case of the ultrarelativistic kinetic ($m=0$) it simplifies
to the {\it quasilinear} form
\begin{equation}
{dp^\prime \over dt^\prime}=-{\gamma \over {\rm sign}(p^\prime) +V}
+ {(1-V^2)^{1/4} \over
1+V {\rm sign}(p^\prime)}  {\xi_{t^\prime}}.
\end{equation}

The certain characteristics of the kinetic (28)
are in principle known from the general Pontryagin-type
equations [16] in terms of two {\it quadratures} of
the stationary probability density distribution and of
the coefficients  of  the  Langevin equation.
We consider the relaxation time ${\tt T}^\prime$ [17]
of the stationary autocorrelation
function (in the moving reference frame)
${\tt T}^\prime = \int_0^\infty dt^\prime
\bigl[\langle p^\prime_{t^\prime} p^\prime_0 \rangle -
\langle  {p^\prime} \rangle^2 \bigr]/
\bigl[\langle  {p^\prime}^2 \rangle
-\langle  {p^\prime} \rangle^2 \bigr]$, which is
a quantity known
from the Jung and Risken formula [18].
Unfortunately,
the integrals cannot be {\it explicitly}
carried out for the  general case (28), so we restrict the
consideration to the
ultrarelativistic kinetic (32) only. Then the calculation is
elementary and  results in
\begin{equation}
{\tt T}^\prime = (\gamma\beta)^{-1} (1-V^2)^{-1/2} (5+8V^2)/(2+V^2).
\end{equation}
Thus, from the viewpoint of a moving observer the decay of the
stationary correlations proceeds slower, comparing to ${\tt T}=5/(2\gamma\beta)$
in the resting reference frame.
Among the usual Lorentz
dilatation some additional correction appears, resulting from the
mentioned difference of the meaning of the averages in
both reference frames.

The general conclusion of the paper is the following.
The {\it invariance principle}, in the context of the thermodynamical
description of a free particles system in the thermal equilibrium, requires
$W^\prime (p^\prime)$ in a moving reference frame to be related with the
Gibbs--Boltzmann
canonical partition function $W(p)$ (in the resting reference frame) via
Eq.~(6). The resulting probability density
distribution (10) is, as the object of the probability theory,
{\it essentially} different from the distribution (4).
The certain relations of the type (19), (21), (24) between quantities
of the
mechanical (mean energy, pressure) and the nonmechanical
(entropy, temperature) origin are established only in a distinguished
resting reference frame, of the considered system as a whole and of the
thermal bath, in which they remain (in contact and) in thermal equilibrium.
The usual identification of thermodynamical functions with
certain averages is not possible in the moving reference frame
description.  The thermodynamical functions in the moving reference frame
results from formulas (12), (20) and (23). Within the Markovian diffusion
theory the kinetic is provided by Eqs.~(5) and (9), or for  the 1D case
{\it explicitly} by Eq.~(27) or (28). The correlation time ${\tt T}^\prime$
for 1D ultrarelativistic case is given by Eq.~(33).

\end{document}